\documentstyle[12pt,fullpage,subeqnarray,setspace]{article}
\def\be{\begin{equation}}
\def\ee{\end{equation}}
\def\beq{\begin{eqnarray}}
\def\eeq{\end{eqnarray}}
\def\bes{\begin{subeqnarray}}
\def\ees{\end{subeqnarray}}
\def\lsim{\:\raisebox{-0.5ex}{$\stackrel{\textstyle<}{\sim}$}\:}

\def\d{\displaystyle}

\begin{document}
\onehalfspace
\begin{flushright}
TIFR/TH/94-50
\end{flushright}
\bigskip
\bigskip
\begin{center}
{\large{\bf Upper Bound on the Mass of Lightest Higgs Boson\\ in the
Supersymmetric Singlet Majoran Model}}\\
\bigskip
\bigskip
{\large{\bf P.N. Pandita}}\\
\bigskip
Theoretical Physics Group \\
Tata Institute of Fundamental Research \\
Homi Bhabha Road, Bombay 400 005, India \\
\bigskip
and \\
\bigskip
Department of Physics \\
North Eastern Hill University \\
Laitumkhrah, Shillong 793 003, India$^\dagger$ \\
\bigskip
\bigskip
{\Large{\bf Abstract}}\\
\end{center}
\smallskip

We calculate an upper bound on the lightest Higgs boson mass in the
supersymmetric singlet majoran model containing $SU(2) \times U(1)$
singlet right handed neutrino fields and a chiral singlet having two
units of lepton number.  We show that although the singlet vacuum
expectation values and supersymmetry breaking masses do not decouple
from the upper bound, even at the tree level, their effect is small,
thereby reducing the bound to the corresponding bound in the Minimal
Supersymmetric Standard Model.  Radiative corrections to the bound are
similar to those in the minimal model.  We also show that the
couplings of the lightest Higgs boson to the majorans are suppressed
in this model.

\vspace{1cm}

\bigskip
\bigskip
\hrule width 5cm
\bigskip

\noindent $^\dagger$ Permanent address

\newpage

Recently considerable attention has been devoted to the study of Higgs
bosons of the Minimal Supersymmetric Standard Model (MSSM) based on
the gauge group $SU(2) \times U(1)$ with two Higgs doublets [1], and
its extensions involving extended Higgs sectors [2,3].  It has been
shown that a calculable upper bound on the mass of the lightest Higgs
boson exists in these models if they remain perturbatively valid below
some scale $\wedge$.  This upper bound can vary between 150 GeV to 175
GeV depending on the Higgs structure of supersymmetric model.  Thus,
nonobservation of a light Higgs boson below this upper bound will rule
out an entire class of Supersymmetric Standard Models with
arbitrary Higgs sectors.

The existence of the upper bound on the lightest Higgs boson mass in
the Minimal Supersymmetric Standard model with arbitrary Higgs sectors
has been investigated in a situation wherein the underlying
supersymmetric model respects the discrete $R$-parity $(R_p)$
symmetry, under which all the standard model particles are even and
their superpartners are odd, so that all renormalizable gauge
invariant baryon $(B)$ and lepton number $(L)$ violating terms in the
superpotential are forbidden.  However, there is no fundamental reason
to believe in exact $R_p$ conservation in the Minimal Supersymmetric
Standard Model.  Indeed $R$-parity would be violated if the lepton
number is not an exact symmetry of nature, as would be the case if
neutrino's have nonvanishing masses, because $R$-parity is related to
the lepton number through $R_p = (-1)^{3B+L+2S}$.  Theoretically, one
can envisage two possibilities of lepton number $(L)$ violation in
supersymmetric theories : either $L$ is broken explicitly by gauge
invariant renormalizable interactions even with the minimal set of
fields as in $MSSM$ [4], or it is an exact global symmetry
spontaneously broken, leading to the existence of a Goldstone boson,
the majoran [5,6].  The second option can, in general, be implemented
only through the introduction of extra fields in supersymmetric
models.  It is also possible to break lepton number spontaneously in
$MSSM$, since the scalar partner of the neutrino can acquire a
nonvanishing vacuum expectation value [7].  In this case the majoran
is dominantly the supersymmetric partner of the neutrino and should be
detected in $Z^0$ decays.  Recent measurements of $Z^0$ width at $LEP$
have ruled out this possibility [8,9].  On the other hand if the
spontaneous breaking of $L$ is induced by the vacuum expectation value
of gauge singlet right-handed neutrino, then the existence of majoran,
which will now be dominantly a gauge singlet object, is not in
contradiction with $Z^0$ width measurements.  Several ways to
implement this idea have been suggested [10,11,12].  The most
economical is the one in which $MSSM$ is extended by including gauge
singlet right-handed neutrino chiral superfields and an additional
gauge singlet superfield, carrying two units of lepton number [12].
This model, the so-called singlet majoran model, has been extensively
discussed recently [12], particularly as regards the spontaneous
breakdown of $R$-parity through radiative effects, as well as the
neutrino properties and related issues.  Since all the singlet majoran
models introduce extra singlet chiral superfields, it is natural to
ask about the existence of an upper bound on the mass of the lightest
Higgs boson in this class of models.  In this paper we shall
investigate the existence of such an upper bound in the simplest of
these models, namely the supersymmetric singlet majoran model [12].
We shall also consider the couplings of the Higgs boson, particularly
the lightest Higgs boson, to majorans in this model.  Both these
questions are of considerable importance in the search for the
lightest Higgs boson in supersymmetric models.

The superpotential of the singlet majoran model which is invariant
under the gauge symmetry $SU(2) \times U(1)_Y$ and the global lepton
number symmetry $U(1)_L$ is ($i,j$ are generation indices) [12]
\beq
W &=& \Big[h_{Uij} H_2 Q_i U^c_j + h_{Dij} H_1 Q_i D^c_j + h_{Lij} H_1
L_i E^c_j \nonumber \\[2mm]
& & ~+ \mu H_1 H_2\Big] + \Big[h_{\nu ij} H_2 L_i N_j + \lambda_{ij} N_i
N_j \Phi\Big],
\eeq
where $Q,L$ are the left-handed quark and lepton doublet superfields,
respectively; $U^c,D^c,E^c$ are the (charge conjugate of) left handed
charge $-2/3$ and $1/3$ quark singlet and lepton superfields,
respectively; $H_1$, $H_2$ are the two Higgs doublets which give
masses to leptons and quarks through the Yukawa couplings
corresponding to the first three terms in the first square bracket of
(1); $N$ are the right-handed neutrino chiral superfields, and $\Phi$
is an additional gauge singlet superfield carrying two units of lepton
number.  The tree level potential for (1) can be written as sum of
three parts,
\be
V_0 = V^0_H (H_1,H_2) + V^0_{N\Phi} (N_i,\Phi) + V^0_\nu
(\nu_i,N_i,\Phi,H_1,H_2),
\ee
where $V^0_H$ is the standard tree level potential of the $MSSM$
[13,~14].  Assuming $CP$ invariance and working in the diagonal basis
for the couplings $\lambda_{ij}$ $(\lambda_{ij} = \lambda_i
\delta_{ij})$ we can write
\beq
V^0_{N\Phi} &=& \sum_i m^2_{N_i} |N_i|^2 + m^2_\Phi |\Phi|^2 -
\left[\sum_i A_i \lambda_i N^2_i \Phi + {\rm h.c.}\right] \nonumber
\\[2mm] & & + 4 \sum_i \bigg|\lambda_i N_i \Phi\bigg|^2 +
\bigg|\sum_i \lambda_i N^2_i\bigg|^2,
\eeq
where $m_{N_i}$, $m_\Phi$ and $A_i$ are the appropriate soft
supersymmetry breaking masses and the trilinear couplings,
respectively.  Since we expect the couplings $h_\nu$ to be of the same
order of magnitude as the Yukawa couplings $h_L$ for charged leptons
$(h_\nu \sim h_L \approx 10^{-2} - 10^{-6})$, we can write to leading
order in $h_\nu$ [15]
\beq
V^0_\nu &=& \sum_i m^2_{\nu_i} |\nu_i|^2 + \d{(g^2 +
g^{\prime 2}) \over 8} \left[\left(\sum_i |\nu_i|^2\right)^2 + 2
\sum_i |\nu_i|^2 \left(|H^0_1|^2 - |H^0_2|^2\right)\right] \nonumber
\\[2mm] & & + \left[\sum_{i,j} h_{\nu_{ij}} \nu_i \left(2\lambda_j
N^\star_j \Phi^\star H^0_2 - \mu N_j H^{0^\star}_1 - A^h_{ij} N_j
H^0_2\right) + {\rm h.c.}\right] \nonumber \\[2mm]
& & + ~{\cal O} (h^2_\nu),
\eeq
where $m_{\nu_i}$ and $A^h_{ij}$ are the appropriate soft
supersymmetry breaking masses and trilinear couplings, respectively,
and $g,\:g'$ are the $SU(2)_L$ and $U(1)_Y$ gauge couplings,
respectively.  Assuming, as in the case of $MSSM$, that all
supersymmetry breaking masses and trilinear couplings are equal to a
universal mass $m_0$ ($\sim 10^2 - 10^3$ GeV) and a universal coupling
$A$, respectively, at some grand unified scale $M_U$ ($\simeq 10^{16}$
GeV), the values of these parameters appearing in the scalar potential
can then be obtained by solving the appropriate renormalization group
equations with boundary conditions at $M_U$.  The effect of running of
these parameters is to drive $SU(2)_L \times U(1)_Y$ breaking through
$V^0_H (H_1,H_2)$ with the nonvanishing vacuum expectation values of
the Higgs doublets $\langle H_1 \rangle = v_1 = v \cos \beta$ and
$\langle H_2 \rangle = v_2 = v \sin \beta$, with $\tan \beta =
v_2/v_1$.  Furthermore, for a wide range of parameters the nontrivial
global minimum of $V^0_{N\Phi} (N_i,\Phi)$ is realized in such a
manner as to break $U(1)_L$ and $R$-parity:
\be
\langle \Phi \rangle = \phi, ~~~\langle N_i \rangle = y_i,
\ee
where $\phi \sim y_i \sim m_0$.  Then, nonzero vacuum expectation
values are induced for the sneutrinos through the $h_\nu$ coupling
which connects the ordinary doublet Higgs and lepton sectors to the
singlet sector:
\be
\langle \nu_i \rangle = x_i \simeq \d{\sum_j h_{\nu ij} y_j
\left[\mu v_1 + (A^h_{ij} - 2\lambda_j \phi)v_2\right] \over
\left[m^2_{\nu_i} + (M^2_Z/2)\cos 2\beta\right]}.
\ee
If $m^2_{\nu_i} \sim m^2_0$, $\phi \sim y_i \sim \mu \sim m_0$, $v_1
\sim v_2 \sim M_W$ and $A^h \sim m_0$, then
\be
x_i \sim h_\nu M_W.
\ee
The spontaneous violation of global lepton number produces a
Nambu-Goldstone boson, the so-called majoran, which can be written as
\be
J = {1 \over \sqrt{2}\: v_R} Im\left[2\phi \Phi - y_i N_i + x_i \nu_i -
\left(\d{\sum_i x^2_i \over v}\right) (\cos \beta H^0_1 -
\sin \beta H^0_2)\right],
\ee
where the lepton-number/$R$-parity violating scale is given by
\be
v_R = \left[4\phi^2 + \sum_i (y^2_i + x^2_i)\right]^{1/2}.
\ee

Starting from the potential (2) it is straightforward to derive the
nondiagonal mass matrix for the scalar Higgs bosons, whose eigenvalues
will provide the masses of physical scalar Higgs particles.  However,
for the specific purpose of the determination of the general bound for
the lightest Higgs boson mass we do not need the full $9 \times 9$
matrix.  All that we need is the upper-left-corner $2 \times 2$
submatrix in any basis, because a known property of any Hermitian
matrix is that its smallest eigenvalue must be smaller than that of
this $2 \times 2$ submatrix.  Denoting by $m^2_{ij}$ the matrix
elements of the squared mass matrix $[m^2]$ for the scalar Higgs
bosons, and by $m^2_{h^0}$ the lightest physical Higgs boson mass, we
obtain the rigorous bound:
\be
m^2_{h^0} \leq {m^2_{11} + m^2_{22} \over 2}\left[1 - \left[1 - 4
\d{m^2_{11} m^2_{22} - m^4_{12} \over (m^2_{11} +
m^2_{22})^2} \right]^{1/2}\right].
\ee
Using the expressions (2), (3) and (4) for the tree level potential
and imposing the minimization conditions $\partial V_0/\partial v_{1,2}
= 0$, we obtain the following expressions for
$m^2_{11},m^2_{22}$ and $m^2_{12}$ in the basis where we have chosen
the $(1,2)$ indices of the mass matrix $[m^2]$ to correspond to
$H^0_1,H^0_2$:
\begin{eqnarray}
m^2_{11} &=& m^2_3 \tan \beta + v^2_1 \left[\d{g^2 +
g^{\prime 2} \over 2}\right] + \d {\mu \over v_1} \sum_{i,j} h_{\nu ij}
x_i x_j, \\
m^2_{22} &=& m^2_3 \cot \beta + v^2_2 \left[\d{g^2 +
g^{\prime 2} \over 2}\right] + \sum_{i,j} h_{\nu ij} \d {x_i y_j \over 2}
(A^h_{ij} - 2\lambda_j \phi), \\
m^2_{12} &=& - m^2_3 - v_1 v_2 \d \left[{g^2 + g^{\prime 2} \over
2}\right],
\end{eqnarray}
where $m^2_3 = B \mu$, with $B$ being the bilinear supersymmetry
breaking parameter, is the standard variable which enters the
potential $V^0_H (H_1,H_2)$ of the $MSSM$.  We note that $m^2_{11}$
and $m^2_{22}$ contain terms which depend on the vacuum expectation
values, $x_i$ and $y_i$, of the singlet fields.  We also note that the
structure of the $2 \times 2$ submatrix (11) -- (13) is different as
compared to the structure of the corresponding $2 \times 2$ submatrix
that occurs in the $MSSM$ with arbitrary Higgs sectors [3].  Using
(11) -- (13) in (10), we get the upper bound
\beq
\d{m^2_{h^0} \over v^2}\; \leq \; \d{1\over2} (g^2 +
g^{\prime 2}) \cos^2 2\beta\!\! &+& \!\! \d{\sum_i}
\d{x^2_i \over v^2} \left[\d{m^2_{\nu_i} \over
v^2} + {1 \over 4} (g^2 + g^{\prime 2}) \left(\cos 2\beta +
\d{\sum_j x^2_j \over v^2}\right)\right] \nonumber \\[2mm]
&+&\!\! \d{\sum_{i,j,k}} \d{x_i x_j \over v^2}
\left[h_{\nu i k} h_{\nu j k} \left(\sin^2 \beta + \d{y^2_k
\over v^2}\right)\right],
\eeq
where we have used the minimization conditions $\partial V_0/\partial
x_i = 0$.  We have made no approximations, retaining all powers of
$x_i$ and $h_\nu$, in obtaining the bound (14), which is, therefore,
exact.

We note that the tree level upper bound (14) depends on supersymmetry
breaking mass parameters $m_{\nu_i}$, as well as singlet vacuum
expectation values $y_i$ and $x_i$.  This is contrary to what happens
in $MSSM$ [13], where the corresponding upper bound is independent of
the supersymmetry breaking masses and is controlled by weak scale
alone.  This is also different from what happens in $MSSM$ with an
extended Higgs sector [16,2,3], where also the bound is controlled by
$v$ alone and is independent of exotic vacuum expectation values (like
those of singlets).  However, we note that the dependence of the bound
(14) on $m^2_{\nu_i}$ and $y_i$ (which can be large and hence make the
bound uncalculable) is controlled by $(x^2_i/v^2)$ which, because of
(7), is a small parameter.  There is a marked hierarchy in the values
of $y_i$ and $x_i$, because $x_i$ is related to the Yukawa coupling
$h_\nu$ through (6) and vanishes as $h_\nu \rightarrow 0$.  The Yukawa
coupling $h_\nu$ is expected to be small $\lsim 10^{-3}$, in order
that stellar energy loss via majoran emitting processes is naturally
suppressed [17,12].  We also note that the dependence of the upper
bound in (14) on $y_i$ is further suppressed by the factor $h^2_\nu$.
Thus, although the singlet vacuum expectation values and the
supersymmetry breaking masses do not decouple in the bound (14), in
practice the (tree) level upper bound on the lightest Higgs mass in
the supersymmetric singlet majoran model is controlled by the weak
scale, and numerically coincides with the corresponding upper bound in
the Minimal Supersymmetric Standard Model.

We now compute the radiative corrections to the tree level bound (14),
using the full one-loop effective potential [18]:
\be
V_1 (Q) = V_0 (Q) + {1 \over 64\pi^2} ~{\rm Str}~M^4\left(\ell n {M^2
\over Q^2} - {3 \over 2}\right),
\ee
where $V_0 (Q)$ is the tree level potential evaluated with couplings
renormalized at some scale $Q$, and where $M^2$ is the field dependent
mass matrix of the particles which contribute to the one-loop
effective potential.  We shall consider only the dominant top-stop and
bottom-sbottom contributions in the one-loop effective potential (15).
For a general sfermion system of two physical states (like stop or
sbottom), we have
\bes
m^2_{1(2)} & = & M^2 \pm \Delta^2, \\[2mm]
M^2 &=& {1\over2} (M^2_{LL} + M^2_{RR}), ~\Delta^2 = {1\over2}
\sqrt{(M^2_{LL} - M^2_{RR})^2 + 4M^2_{LR}}, \\[2mm]
M^2_{LL} &=& m^2_{{\rm soft},L} + m^2_f - \left[{g^2 \over 2} T_{3L} -
{g' \over 2} Y_L\right] \left(v^2_2 - v^2_1 - \sum_i x^2_i\right), \\[2mm]
M^2_{RR}& = & M^2_{LL} ~~{\rm with}~~ (L \Leftrightarrow R),
\ees
where $m^2_{{\rm soft},L(R)}$ are the squared $SUSY$-breaking masses
which, in general, are not equal at the electroweak scale, $m_f$ is
the mass of the corresponding fermion, and $T_{3L,3R}$ and $Y_{L,R}$
are the $SU(2)_L$ and $U(1)_Y$ charges of the left and right
sfermions.  The mixing term $M^2_{LR}$ of the sfermion mass matrix is
given by
\bes
M^2_{LR} &=& -h_t \left(A_t v_2 + \mu v_1 - \sum_{i,j} h_{\nu ij} x_i
y_j\right), ~~~{\rm (for~stop)}, \\[2mm]
&=& h_b (A_b v_1 + \mu v_2), ~~~{\rm (for~sbottom)},
\ees
respectively.  Using the results (16) and (17) in (15), we obtain an
expression for the mass matrix at one loop that contains the
contribution of all field-dependent masses.  For the radiatively
corrected upper bound on the lightest Higgs mass, we shall be
interested in the upper-left-corner $2 \times 2$ submatrix of this
mass matrix.  Writing the matrix elements of this submatrix as
\be
m^2_{ij} = m^{(0)^2}_{ij} + \delta m^2_{ij}, ~~~i,j = 1,2,
\ee
where $m^{(0)^2}_{ij}$ are formally analogous to the tree-level
expressions (11) -- (13) which now depend on the renormalized couplings
evaluated at scale $Q^2$ ($= M^2_Z$ in the present case), we obtain
the following expressions for the fermion-sfermion contributions to
the quantities $\delta m^2_{ij}$:
\bes
\delta m^2_{11} &=& \left\{- \d{3 \over 64 \pi^2} \d{g^2 \over
\sin^2\beta} \d{m^2_t \over m^2_W} \d{A_t \mu \over (m^2_{t^\sim_1} -
m^2_{t^\sim_2})} \left[f(m^2_{t^\sim_1}) - f(m^2_{t^\sim_2})\right] +
(t \rightarrow b)\right\} \tan \beta \nonumber \\[2mm]
& & ~+ ~\left\{\d{3 \over 64\pi^2} \d{g^2 \over \sin^2 \beta} \d{m^2_t
\over m^2_W} \d{\mu \over (m^2_{t^\sim_1} - m^2_{t^\sim_2})}
\d{\sum_{i,j} h_{\nu ij} x_i y_j \over v_1} \left[f(m^2_{t^\sim_1}) -
f(m^2_{t^\sim_2})\right]\right\} \nonumber \\[2mm]
& & ~+ \left\{\d{3 \over 16\pi^2} \d{g^2 \over m^2_W}
\Delta_{11}\right\}, \\[2mm]
\delta m^2_{22} &=& \left\{-\d{3 \over 64\pi^2} \d{g^2 \over \sin^2
\beta} \d{m^2_t \over m^2_W} \d{A_t \mu \over (m^2_{t^\sim_1} -
m^2_{t^\sim_2})} \left[f(m^2_{t^\sim_1}) - f(m^2_{t^\sim_2})\right] +
(t \rightarrow b)\right\} \cot \beta \nonumber \\[2mm]
& &  ~+ \left\{\d{3 \over 64\pi^2} \d{g^2 \over \sin^2\beta} \d{m^2_t
\over m^2_W} \d{A_t \over (m^2_{t^\sim_1} - m^2_{t^\sim_2})}
\d{\sum_{i,j} h_{\nu i j} x_i y_j \over v_2} \left[f(m^2_{t^\sim_1}) -
f(m^2_{t^\sim_2})\right]\right\} \nonumber \\[2mm]
& &  ~+ \left\{\d{3 \over 16\pi^2} \d{g^2 \over m^2_W}
\Delta_{22}\right\}, \\[2mm]
\delta m^2_{12} &=& \left\{\d{3 \over 64\pi^2} \d{g^2 \over \sin^2
\beta} \d{m^2_t \over m^2_W} \d{A_t \over (m^2_{t^\sim_1} -
m^2_{t^\sim_2})} \left[f(m^2_{t^\sim_1}) - f(m^2_{t^\sim_2})\right] +
(t \rightarrow b)\right\} \nonumber \\[2mm]
& &  ~+ \left\{\d{3 \over 16\pi^2} \d{g^2 \over m^2_W}
\Delta_{12}\right\},
\ees
where
\bes
\Delta_{11} &=& \d{m^4_t \over \sin^2 \beta} \left[\mu^2 A^{\prime
2}_T \; g(m^2_{t^\sim_1},m^2_{t^\sim_2})\right] \nonumber \\[2mm]
& &  ~+ \d{m^4_b \over \cos^2 \beta} \left[\ell n\left(\d{m^2_{b^\sim_1}
m^2_{b^\sim_2} \over m^4b}\right) + 2A_b A_B \ell
n\left({m^2_{b^\sim_1} \over m^2_{b^\sim_2}}\right) + A^2_b A^2_B\:
g(m^2_{b^\sim_1},m^2_{b^\sim_2})\right], \\[2mm]
\Delta_{22} &=& {m^4_t \over \sin^2\beta} \left[\ln
\left({m^2_{t^\sim_1} m^2_{t^\sim_2} \over m^4_t}\right)
+ 2A_tA'_T \ln \left({m^2_{t^\sim_1}
\over m^2_{t^\sim_2}}\right) + A^2_t A'^2_T\: g\!\left(m^2_{t^\sim_1},
m^2_{t^\sim_2}\right)\right] \nonumber \\[2mm]
& + & {m^4_b \over \cos^2 \beta} \left[\mu^2 A^2_B\:
g\!\left(m^2_{b^\sim_1}, m^2_{b^\sim_2}\right) \right] \\[2mm]
\Delta_{12} &=& \d{m^4_t \over \sin^2 \beta} \mu A'_T \left[\ell
n\left(\d{m^2_{t^\sim_1} \over m^2_{t^\sim_2}}\right) + A_t A'_T \;
g\left\!(m^2_{t^\sim_1},m^2_{t^\sim_2}\right)\right] \nonumber \\[2mm]
& &  ~+ \d{m^4_b \over \cos^2 \beta} \mu A_B \left[\ell
n\left(\d{m^2_{b^\sim_1} \over m^2_{b^\sim_2}}\right) + A_b A_B\;
g\left\!(m^2_{b^\sim_1},m^2_{b^\sim_2}\right)\right].
\ees
The functions $f(m^2)$, $g(m^2_1,m^2_2)$, $A'_T$ and $A_B$ are defined
as
\bes
f(m^2) &=& 2m^2\left[\ell n\left(\d{m^2 \over Q^2}\right) - 1\right],
{}~g(m^2_1,m^2_2) = 2 - \left(\d{m^2_1 + m^2_2 \over m^2_1 -
m^2_2}\right) \ell n\left(\d{m^2_1 \over m^2_2}\right), \\[2mm]
A'_T &=& A_T - \sum_{i,j} \d{h_{\nu ij} x_i y_j \over v_2}, ~A_T =
\d{A_t + \mu \cot \beta \over (m^2_{t^\sim_1} - m^2_{t^\sim_2})}, ~A_B
= \d{A_b + \mu \tan \beta \over (m^2_{b^\sim_1} - m^2_{b^\sim_2})}.
\ees
We note that in obtaining the results (18) -- (21), we have neglected
the contribution of $D$-terms in the squark masses (16) since we are
including only the quark-squark contributions to $V_1 (Q)$.  We
further note that the radiative corrections depend on the singlet
vacuum expectation values $x_i$ and $y_i$.

An absolute upper bound on the radiatively corrected mass of the
lightest Higgs mass can be obtained by finding the smaller eigenvalue
of the $2 \times 2$ submatrix (18).  We then obtain the upper bound
\beq
{m^2_{h^0} \over v^2} &\leq& \d{1\over2} (g^2 + g^{\prime 2}) \cos^2 2\beta +
\sum_i \d{x^2_i \over v^2} \left[\d{m^2_{\nu_i} \over v^2} +
\d{1\over4} (g^2 + g^{\prime 2}) \left(\cos 2\beta + \d{\sum_j x^2_j
\over v^2}\right)\right] \nonumber \\[2mm]
& & + \sum_{i,j,k} \d{x_i x_j \over v^2} \left[h_{\nu i k} h_{\nu j k}
\left(\sin^2 \beta + \d{y^2_k \over v^2}\right)\right] \nonumber
\\[2mm] & & + \d{3 \over 64\pi^2} \left(\d{g^2 m^2_t \over m^2_W v
\sin \beta}\right) \d{\left[f(m^2_{t^\sim_1}) -
f(m^2_{t^\sim_2})\right] \left[A_t + \mu \cot \beta\right] \over
m^2_{t^\sim_1} - m^2_{t^\sim_2}} \sum_{i,j} h_{\nu ij} x_i y_j
\nonumber \\[2mm]
& & + \d{3 \over 16\pi^2} \left(\d{g^2 \over m^2_W}\right)
\left[\Delta_{11} \cos^2 \beta + \Delta_{22} \sin^2 \beta +
\Delta_{12} \sin 2\beta\right],
\eeq
to be compared with the tree level bound (14).  We note that this
bound is independent of $\lambda_i$.  On the other hand the bound
depends quadratically on the singlet vacuum expectation values $x_i$
and $y_i$.  Thus the one loop bound (22) is, in principle, not
screened from the large vacuum expectation value $y_i$.  This is in
contrast to what happens in the nonminimal supersymmetric standard
model [19], where the radiative corrections to the lightest Higgs mass
depend only logarithmically on the singlet vacuum expectation value
and are, thus, screened from the lightest Higgs mass bound.  On the
other hand the dependence on $y_i$ is controlled by $h_\nu$, which is
a small number, and hence numerically unimportant.  We also note that
the functions $\Delta_{ij}$ $(i,j = 1,2)$ in (20) are numerically very
close to the corresponding functions in $MSSM$ [1] as well the
nonminimal supersymmetric standard model [2], because the deviations
from $MSSM$ are given by $h_{\nu ij} x_i y_i/v_2 \lsim 10^{-3}$ GeV
for $h_\nu \sim 10^{-3}$, $x_i \sim 10^{-1}$ GeV, and $y_i \sim 10^3$
GeV, and are insignificant.  Thus, in practice the upper bound (22) is
numerically the same as in the Minimal Supersymmetric Standard Model.

We now turn our attention to the couplings of the lightest Higgs boson
in the singlet majoran model.  We shall be particularly interested in
the couplings to the majorans.  Writing the weak scalar and
pseudoscalar fields as a vector $Z_i \equiv (H_1,H_2,\nu_i,\Phi,N_i)$,
and shifting the fields $Z_i = v_i + (Re~Z_i + Im~Z_i)/\sqrt{2}$, we
can define the mass eigenstates by $h_i = P_{ij} x_j$, $A_i = Q_{ij}
y_j$, where $P$ and $Q$ are orthogonal matrices.  The indices $i,j =
1,2,\cdots,9$, and denote the scalar and pseudoscalar bosons involved
($h_1 \equiv h^0$, the lightest Higgs boson), ordered according to
their increasing masses.  Substituting in (2) and keeping only the
leading terms in (8) we get the Higgs-majoran lagrangian
\beq
{\cal L}_{h_i JJ} &=& - \Bigg \{\sqrt{2} P_{ik} y_k \left[2\lambda^2_k -
4 A_k \lambda_k \d{\phi \over v^2_R} + 8 \d{\lambda^2_k \over
v^2_R}\right] \nonumber \\[2mm]
& & ~~+\sqrt{2} P_{i6} \phi \left[4\d{y^2_k \lambda^2_k} + \d{A_k
\lambda_k y^2_k \over \phi v^2_R}\right]
+ 4\sqrt{2} h_{\nu jk} \lambda_k (x_j P_{i2} + v_2 P_{ij})
\left(\d{\phi y_k \over v^2_R}\right)
\nonumber \\[2mm]
& & ~~ + \left[h_{\nu jk} h_{\nu lk}
(x; P_{il} + x_l P_{ij}) + 2 h_{\nu lk} h_{\nu lk} (\nu_2 P_{iz})
\right]  \left({y^2k \over \sqrt 2\:\nu^2_R}\right)
\Bigg\} (h_i JJ),
\eeq
where repeated indices are summed.  We see from (23) that the
couplings of the lightest higgs $h_1 ~(\equiv h^0)$ to the majorans is
suppressed unlike what happens in other supersymmetric majoran models,
where such couplings are unsuppressed [11].  Thus the dominant decay
modes of the lightest Higgs boson in the supersymmetric singlet
majoran model are essentially same as in the $MSSM$.

In conclusion we have obtained an upper bound on the lightest Higgs
boson mass in the supersymmetric singlet majoran model, and shown that
although the bound depends on the singlet vacuum expectation values
(and at tree level on the supersymmetry breaking masses), the
dependence is weak and the bound coincides with the corresponding
bound in $MSSM$.  We have also shown that the couplings of the
lightest Higgs boson to the majoran is suppressed, and, therefore, the
dominant decay modes are the same as in $MSSM$.  Thus for all
practical purposes the lightest Higgs boson behaves in the same manner as
the lightest Higgs of the $MSSM$.
\bigskip

\noindent {\large{\bf Acknowledgements}} \\
\smallskip

\nobreak
I would like to thank Tata Institute of Fundamental Research for its
hospitality extended to me as an Indian National Science Academy
Visiting Fellow while this work was completed.  This work was
supported by the Department of Atomic Energy, India.

\newpage

\begin{center}
{\Large{\bf References}}\\
\end{center}

\bigskip
\bigskip

\begin{enumerate}

\item[{[1]}] Y. Okada, M. Yamaguchi and T. Yanagida, Prog. Theor.
Phys. {\bf 85}, 1 (1991); Phys. Lett. {\bf B262}, 54 (1991); J. Ellis, G.
Ridolfi and F. Zwirner, Phys. Lett. {\bf B257}, 83 (1991); H.E. Haber
and R. Hempfling, Phys. Rev. Lett. {\bf 66}, 1815 (1991); R. Barbieri,
M. Frigeni and F. Caravaglios, Phys. Lett. {\bf B258}, 167 (1991).
\hfill\break For review and a list of references, see, e.g., P.N.
Pandita, Pramana, Journal of Physics Suppl. {\bf 41}, 303 (1993); H.E.
Haber, in ``Perspectives on Higgs Physics'', Ed. G.L. Kane (World
Scientific, Singapore, 1992).

\item[{[2]}] P.N. Pandita, Phys. Lett. {\bf B318}, 388 (1993); Z. Phys.
{\bf C59}, 575 (1993); T. Elliot, S.F. King and P.L. White, Phys. Lett.
{\bf B305}, 71 (1993); U. Ellwanger, Phys. Lett. {\bf B303}, 271 (1993);
D. Comelli and C. Verzegnassi, Phys. Rev. {\bf D47}, R764 (1993); J.R.
Espinosa and M. Quiros, Phys. Lett. {\bf B279}, 92 (1992); P. Binetruy
and C.A. Savoy, Phys. Lett. {\bf B277}, 453 (1992).

\item[{[3]}] G.L. Kane, C. Kolda and J.D. Wells, Phys. Rev. Lett.
{\bf 70}, 2686 (1993); J.R. Espinosa and M. Quiros, Phys. Lett.
{\bf B302}, 51 (1993).

\item[{[4]}] L.J. Hall and M. Suzuki, Nucl. Phys. {\bf B231}, 419
(1984); S. Dawson, Nucl. Phys. {\bf B261}, 297 (1985); I.H. Lee, Nucl.
Phys. {\bf B246}, 120 (1984); V. Barger, G.F. Guidice and T. Han, Phys.
Rev. {\bf D40}, 2987 (1989); S. Dimopoulos et al., Phys. Rev. {\bf D41},
2099 (1990); K. Enqvist, A. Masiero and A. Riotto, Nucl. Phys.,
{\bf B373}, 95 (1992).

\item[{[5]}] G.B. Gelmini and M. Roncadelli, Phys. Lett. {\bf B99}, 411
(1981).

\item[{[6]}] Y. Chikashige, R.N. Mohapatra and R.D. Peccei, Phys.
Lett. {\bf B98}, 265 (1981).

\item[{[7]}] C.S. Aulakh and R.N. Mohapatra, Phys. Lett. {\bf B119}, 136
(1982); ibid {\bf B121}, 147 (1983); J. Ellis et al., Phys. Lett.
{\bf B150}, 142 (1985); G.G. Ross and J.W.F. Valle, Phys. Lett.
{\bf B151}, 375 (1985).

\item[{[8]}] F. Dydak, in ``Proceedings of XXV High Energy Physics
Conference'', Singapore (1990).

\item[{[9]}] See, however, D. Comelli, A. Masiero, M. Pietroni and A.
Riotto, Phys. Lett. {\bf B324}, 397 (1994).

\item[{[10]}] M.C. Gonzalez-Garcia and J.W.F. Valle, Nucl. Phys.
{\bf B355}, 330 (1991); J.W.F. Valle, Phys. Lett. {\bf B196}, 157 (1987).

\item[{[11]}] A. Masiero and J.W.F. Valle, Phys. Lett. {\bf B251}, 273
(1990); J.C. Rom\~ao, C.A. Santos and J.W.F. Valle, Phys. Lett.
{\bf B288}, 311 (1992); J.C. Rom\~ao, F.de Campos and J.W.F. Valle,
Phys. Lett. {\bf B292}, 329 (1992).

\item[{[12]}] G.F. Giudice, A. Masiero, M. Pietroni and A. Riotto,
Nucl. Phys. {\bf B396}, 243 (1993); M. Shiraishi, I. Umemura and K.
Yamamoto, Phys. Lett. {\bf B313}, 89 (1993); I. Umemura and K. Yamomoto,
Nucl. Phys. {\bf B423}, 405 (1994).

\item[{[13]}] K. Inoue, A. Kakuto, H. Komatsu and S. Takeshita, Prog.
Theor. Phys. {\bf 68}, 927 (1982); R.A. Flores and M. Sher, Ann. Phys.
{\bf 148}, 95 (1983).

\item[{[14]}] We denote the scalar component of a chiral superfield by
the same symbol as the superfield.  In particular $\nu_i$ denotes the
scalar partner of the left handed neutrino.

\item[{[15]}] In actual calculations we shall not make this
approximation, but shall retain all powers of $h_\nu$.

\item[{[16]}] M. Drees, Int. Journal of Modern Physics {\bf A4}, 3635
(1989); J.R. Espinosa and M. Quiros, Phys. Lett. {\bf B279}, 92 (1992).

\item[{[17]}] G.G. Raffelt, Phys. Rep. {\bf 198}, 1 (1990).

\item[{[18]}] S. Coleman and E.J. Weinberg, Phys. Rev. {\bf D7}, 1888
(1973).

\item[{[19]}] Pandita, Ref. 2; Comelli and Verzegnassi, Ref. 2.

\end{enumerate}

\end{document}